\documentclass[trackchanges]{aastex7} 

\usepackage{amsmath}
\usepackage{amsfonts}
\usepackage{amssymb}
\usepackage{mathrsfs}
\usepackage{multirow}
\usepackage{booktabs}
\usepackage{comment}
\usepackage{hyperref}

\begin{document}

\title{Hunting for ``Oddballs'' with Machine Learning:  
Detecting Anomalous Exoplanets Using a Deep-Learned Low-Dimensional Representation of Transit Spectra with Autoencoders}

\author[0000-0003-2719-221X]{Alexander Roman}
\affiliation{Department of Physics and Astronomy, University of Alabama, Tuscaloosa, AL, 35487, USA}
\email[show]{adroman@ua.edu}  

\author[0009-0000-8050-5348]{Emilie Panek}
\affiliation{Department of Physics and Astronomy, University of Alabama, Tuscaloosa, AL, 35487, USA}
\email{epanek1@ua.edu}

\author[0000-0002-0355-2076]{Roy T.~Forestano}
\affiliation{Physics Department, University of Florida, Gainesville, FL 32653, USA}
\email{roy@example.com}

\author[0000-0002-6683-6463]{Eyup B. Unlu}
\affiliation{Physics Department, University of Florida, Gainesville, FL 32653, USA}
\email{eyup@example.com}

\author[0000-0003-3074-998X]{Katia Matcheva} 
\affiliation{Department of Physics and Astronomy, University of Alabama, Tuscaloosa, AL, 35487, USA}
\email{kmatcheva@ua.edu}

\author[0000-0003-4182-9096]{Konstantin T.~Matchev}
\affiliation{Department of Physics and Astronomy, University of Alabama, Tuscaloosa, AL, 35487, USA}
\email{kmatchev@ua.edu}


\begin{abstract}

This study explores the application of autoencoder-based machine learning techniques for anomaly detection to identify exoplanet atmospheres with unconventional chemical signatures using a low-dimensional data representation. We use the Atmospheric Big Challenge (ABC) database, a publicly available dataset with over 100,000 simulated exoplanet spectra, to construct an anomaly detection scenario by defining CO$_2$-rich atmospheres as anomalies and CO$_2$-poor atmospheres as the normal class.  We benchmarked four different anomaly detection strategies: Autoencoder Reconstruction Loss, One-Class Support Vector Machine (1 class-SVM), K-means Clustering, and Local Outlier Factor (LOF). Each method was evaluated in both the original spectral space and the autoencoder’s latent space using Receiver Operating Characteristic (ROC) curves and Area Under the Curve (AUC) metrics. To test the performance  of the different methods under realistic conditions, we introduced Gaussian noise levels ranging from 10 to 50 ppm. Our results indicate that anomaly detection is consistently more effective when performed within the latent space across all noise levels. 
Specifically, K-means clustering in the latent space emerged as a stable and high-performing method. We demonstrate that this anomaly detection approach  is robust to noise levels up to 30 ppm (consistent with realistic space-based observations) and remains viable even at 50 ppm when leveraging latent space representations. On the other hand, the performance of the anomaly detection methods applied directly in the raw spectral space degrades significantly with increasing the level of noise.
This suggests that autoencoder-driven dimensionality reduction offers a robust methodology for flagging chemically anomalous targets in large-scale surveys where exhaustive retrievals are computationally prohibitive. 

\end{abstract}


\keywords{Machine learning --- Anomaly detection --- Autoencoder --- Exoplanet --- Transmission spectroscopy}

\section{Introduction}

In the last three decades the number of discovered extrasolar system planets has grown into the thousands (see \href{https://exoplanetarchive.ipac.caltech.edu/}{https://exoplanetarchive.ipac.caltech.edu/} for the up-to-date number of confirmed exoplanets). The observational efforts have now moved from simple detection of individual planets to launching large planetary surveys that yield hundreds and even thousands of new planets \citep{Borucki_2010,Koch_2010,Ricker_2015,Ricker_2022,Borucki_2016,Borucki_2017,Barclay_2018}. Furthermore, advances in the observational techniques have allowed us to move from detection to actual characterization of the planetary environment revealing exotic conditions that differ starkly from our own Solar system examples. In this process, transit spectroscopy has emerged as an indispensable tool for studying the structure and chemical composition of the atmospheres of extrasolar system planets \citep{Charbonneau_2000, Seager_2000, Burrows_2014}. Coupled with advances in instrumentation and data reduction techniques and the development of ever more detailed numerical models, the field of studying exoplanetary atmospheres is ready for prime time. With more than 600 individual transmission spectra listed on the NASA Exoplanet archive (\href{https://exoplanetarchive.ipac.caltech.edu/index.html}{https://exoplanetarchive.ipac.caltech.edu/index.html}) and the increasing number of high-quality observations using the JWST facilities, analyzing these spectra is becoming more demanding in terms of time and computational resources. Future observatories such as the ESA Ariel mission \citep{Tinetti_2018_ariel, tinetti_2021_ariel} are planning to take these numbers to the extreme and observe 1000 transiting planets in transmission and/or emission. The challenge is on several fronts \citep{Barstow_2020}: 1) the sheer number of observations presents a computational challenge for  atmospheric retrievals; 2) the increased spectral resolution and range of modern instruments, while desirable, results in long processing times; 3) using more complex, realistic atmospheric models significantly increases the time required for numerical analysis. In addition, it is often desired to repeat the data analysis under different assumptions which directly multiplies the amount of time needed to perform a comprehensive data analysis and atmospheric retrievals. 

This motivates the development of alternative methods which would allow for fast, robust, and accurate analysis of large data sets that can be used either as stand alone tools or as a complimentary approach to the classical methods. A number of previous studies have demonstrated the efficiency of Machine Learning (ML) methods as an alternative to sampling-based retrieval models for exoplanetary characterization. These include surrogate ML forward radiative transfer models \citep{Himes_2022}, ML retrieval models that are trained on actual Bayesian retrievals \citep{Eyup_2023}, as well as ML regression models trained on a sampled database. Explored architectures include Bayesian neural networks \citep{Cobb2019}, generative adversarial networks (GANs) \citep{Zingales_2018}, convolutional neural networks (CNNs) \citep{Soboczenski2018,Yip_2021,Ardevol2022,Haldemann2022}, random forests \citep{Marquez2018,Fisher2020,Guzman2020,Nixon2020,2025arXiv250804982F}, normalizing flows \citep{Yip2022,2023arXiv230909337A,2025IEEEA..13m7773O}, symbolic regression \cite{Matchev_2022}, simulation based inference \citep{Lueber_2025}, etc., all of which rely on supervised or semi-supervised methods. Unsupervised ML techniques have also been used as a way of studying the statistical properties of the spectra \citep{2022PSJ.....3..205M} or to provide a guided prior to the standard Bayesian retrievals \citep{Hayes2020}.      

ML techniques are particularly successful at finding correlations and patterns in high dimensional data which can be exploited to answer more targeted questions. \cite{Forestano_2023} demonstrated the use of  anomaly detection techniques to search for unusual planetary spectra, which could be particularly useful in large surveys as a fast approach to mark planets that do not conform to the theoretical expectations and could be of interest for followup observations. The study compared the performance of two popular anomaly detection algorithms: One class support vector machine (1-class SVM) and Local Outlier Factor (LOF) both used as novelty detection methods applied directly on the original planetary spectra. While the methods performed very well on data with low level of noise, their performance gradually deteriorated as the level of noise increased from 10 ppm to 50 ppm.

In this paper we extend the work by \cite{Forestano_2023} and investigate the performance of the ML novelty detection techniques in conjunction with dimensionality reduction methods. The purpose of the dimensionality reduction is to reduce the high-dimensional spectral data to an equivalent low-dimensional representation (referred to as ``latent space'') without any loss of information. This is possible because the number of input model parameters is typically much smaller than the number of spectral bins. The dimensionality reduction thus leverages the existing correlations among the spectral bins and renders the data more amenable to subsequent numerical manipulations. 

We demonstrate that the application of the anomaly detection methods within the lower-dimensional latent space results in improved performance. This has been previously noted in the case of transit light curves \citep{hönes2021_automatically_detecting_anomalous_exoplanet}. We find that the benefit is most notable in cases of higher level of noise where the detection of the anomalies is more challenging. Therefore, this is a promising approach when searching for weak signals in noisy observations, e.g., trace gases or potential biosignatures.

The paper is organized as follows.
In Section \ref{sec:dataset} we introduce the database for our analysis. Section~\ref{sec:ABCoriginal} describes the original ABC database, while Section~\ref{sec:preprocessing} describes the preselection cuts and the subsequent preprocessing steps of the spectra. Section \ref{sec:methods} explains the dimensionality reduction procedure and the implementation of autoencoders as a dimensionality reduction technique. We then introduce several commonly used novelty detection methods (1-class SVM, LOF, and K-means) and compare their performance when applied on the original (spectral) space and on the dimensionally reduced (latent) space. In Section~\ref{sec:results} we summarize the results and conclude in Section \ref{sec:conclusions}.

\section{Dataset description} \label{sec:dataset}

\subsection{The original ABC database}
\label{sec:ABCoriginal}

Machine learning models are only as good as the data they train on. Thus it is prudent to invest extra effort and time to curate a good quality database that is well understood and tested. For our study we will use the Atmospheric Big Challenge (ABC) database \citep{changeat_2023_ABC}, created for the Ariel Data Challenge 2022 \citep{Yip_competition}, with some modifications as described in \cite{Forestano_2023}.  

The ABC database provides 105,887 synthetic transmission spectra, based on a diverse list of 5900 unique exoplanets. The planets were selected from the Ariel preliminary target list comprised from confirmed exoplanets and TESS planetary candidates \citep{Edwards_2022}. While some of the objects might be false positive, they are still representative of the planet population expected to be observed with Ariel. The original spectra reflect the observational capabilities of the Ariel instruments \citep{tinetti_2021_ariel}, with 52 spectral bins covering between 0.5 and 7.5 microns, with each spectrum satisfying the criteria for Ariel tier 2 observations. The spectra were calculated using Alfnoor \citep{Changeat_2020}, which combines the TauREx software \citep{Al_Refaie_2021} for the simulation, and ArielRad \citep{Mugnai_2020} for the binning and the addition of Ariel-type instrumental noise to the simulation. 

For each planet, the stellar and planet physical parameters were fixed to their literature values, while the chemical composition of the atmosphere was randomly generated. We consider planets with H$_2$ and He dominated atmospheres with a He/H$_2$ ratio of 0.17. In addition, trace gases are considered, namely H$_2$O, CH$_4$, CO, CO$_2$, and NH$_3$. The abundances of the trace gases are randomly sampled on a log scale within the ranges shown in Table~\ref{tab:abundances}.

\begin{table}[h]
\centering
 \caption{Sampling ranges for the main absorbers in the ABC database. }
 \label{tab:abundances}
\begin{tabular}{|c|c|}
\hline
Trace gases & Concentration ranges \\
\hline
H$_2$O  & $[10^{-9} ~;~ 10^{-3}]$ \\
CO & $[10^{-6} ~;~ 10^{-3}]$\\
CO$_2$ &$[10^{-9} ~;~ 10^{-4}]$ \\
CH$_4$ & $[10^{-9} ~;~ 10^{-3}]$\\
NH$_3$ &$[10^{-9} ~;~ 10^{-4}]$ \\
\hline
\end{tabular}
\end{table}


The simulated atmospheres also include Collision-Induced Absorption (CIA) for H$_2$-H$_2$ and H$_2$-He as well as Rayleigh scattering. The simulations do not contain any clouds or hazes. The atmospheres were considered to be isothermal and the temperature was fixed at the equilibrium value for the respective planet. The ABC dataset underwent extensive testing and validation during the 2022 Ariel Machine Learning Data Challenge, by members of both the organizing team and the competition teams \citep{yip_2023_pmlr}.  

\subsection{Modified ABC database and data preprocessing}
\label{sec:preprocessing}

The ABC database is well suited for our study because it includes realistic simulations of exoplanet atmospheres based on real targets from the Ariel mission list. It contains a large number of spectra with a wide range of physical properties. However, one of our goals in this paper is to study the performance of different ML methods with increasing levels of noise (we shall use a default noise level of $\sigma = 30$ ppm, but will also consider $\sigma = 10$, $\sigma = 20$, and $\sigma = 50$ ppm). Since the noise-free TauREx spectra were not included in the original ABC database, \cite{Forestano_2023} used TauREx to recalculate the corresponding noiseless spectra based on the same stellar and planetary parameters. In what follows, we shall use this modified noiseless version of the ABC database as our starting point and add the corresponding level of instrumental noise, for which we take a standard Gaussian distribution.

For the purpose of our numerical exercise we performed the following preprocessing steps. First, we removed planets with radii smaller than 1.5R$_{\oplus}$, as they would require different assumptions for their atmospheres (for example these atmospheres would not be dominated by hydrogen and helium). Second, to ensure that the spectra are informative, we applied a quality filter based on the feature height, whereby any spectrum with total variation below 210~ppm is discarded. This threshold reflects an approximate signal-to-noise ratio (SNR) of 7, assuming a noise floor of 30~ppm, and ensures that we focus on spectra with detectable features.

\begin{figure}[h]
    \centering
    \includegraphics[width=\textwidth]{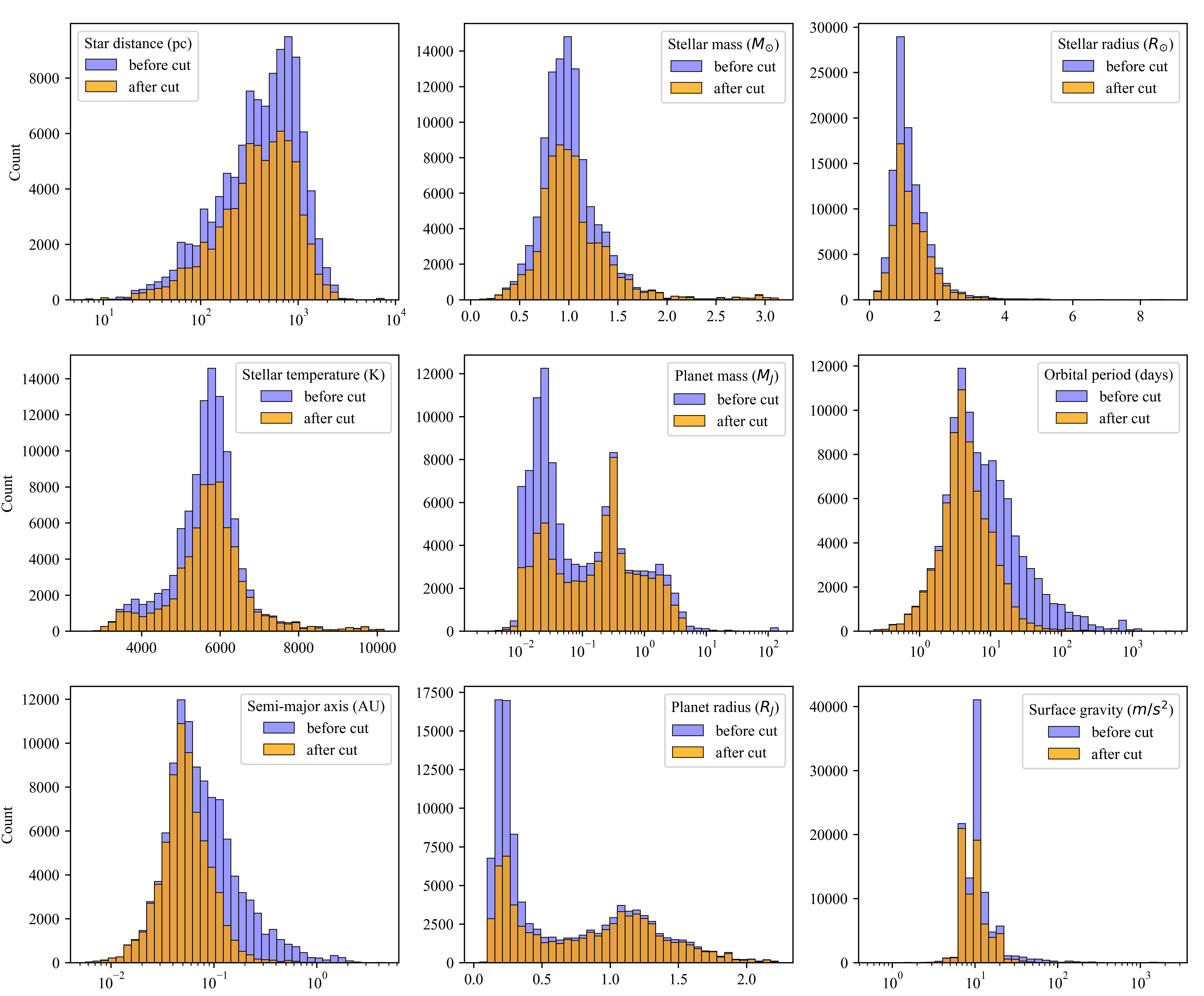}
    \caption{Distributions of selected parameters in the database used in this study (after \cite{Forestano_2023}). The blue histograms represent the distributions over the full ABC database, while the orange histograms represent the distributions within the subset left after the cuts discussed in the text.}
    \label{fig:histogramms_databases}
\end{figure}

After applying these filters, we get a final database of 69,099 spectra. The distributions of parameters before and after these cuts are presented in Figure \ref{fig:histogramms_databases} with blue and orange histograms, respectively. 

\begin{figure}[ht]
    \centering
    \includegraphics[width=\textwidth]{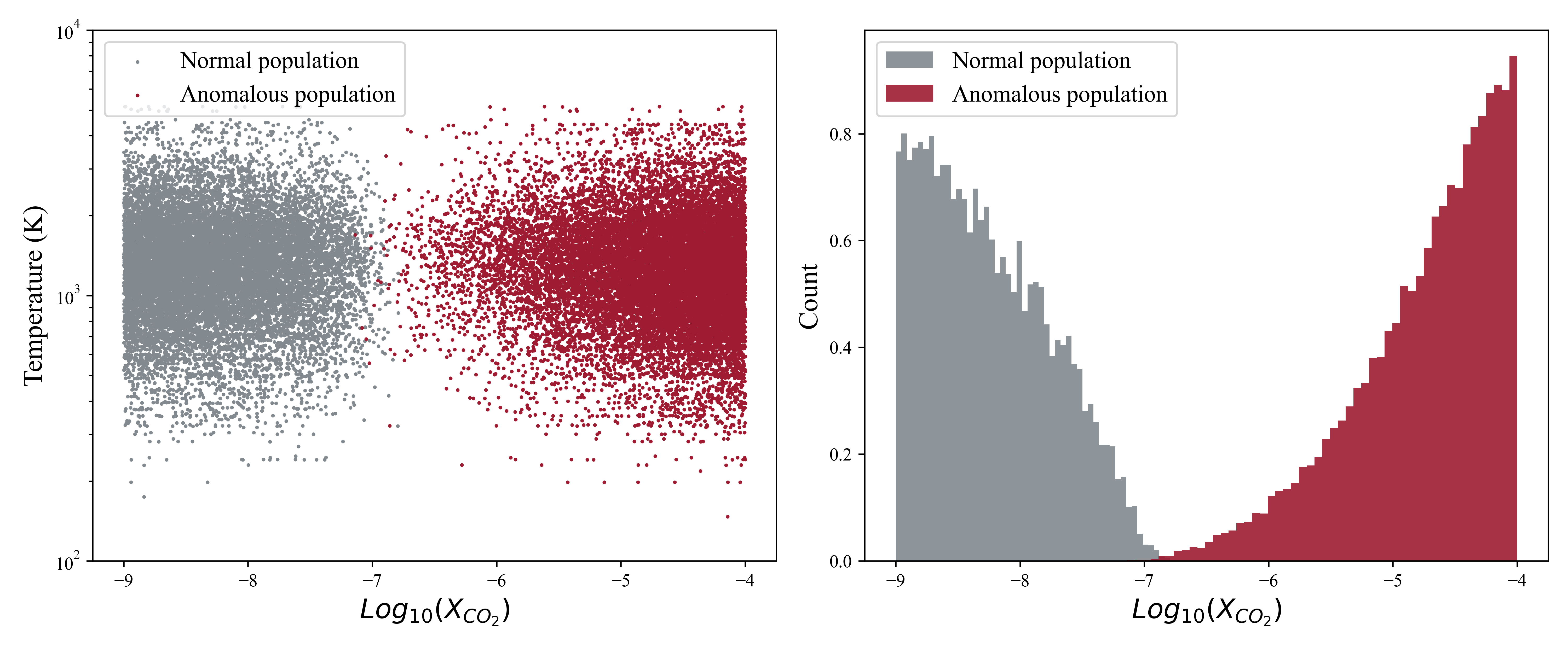}
    \caption{Left panel: Scatter plot of the temperature versus the log concentration of CO$_2$ (after \cite{Forestano_2023}). The gray points represent the normal population and the red points represent the anomalous population. Right panel: Histogrammed distributions of planets according to their log-concentration of CO$_2$: the normal population (gray histogram) and the anomalous population (red histogram).}
    \label{fig:normal_vs_anomalous}
\end{figure}

\begin{figure}
    \centering
    \includegraphics[width=\textwidth]{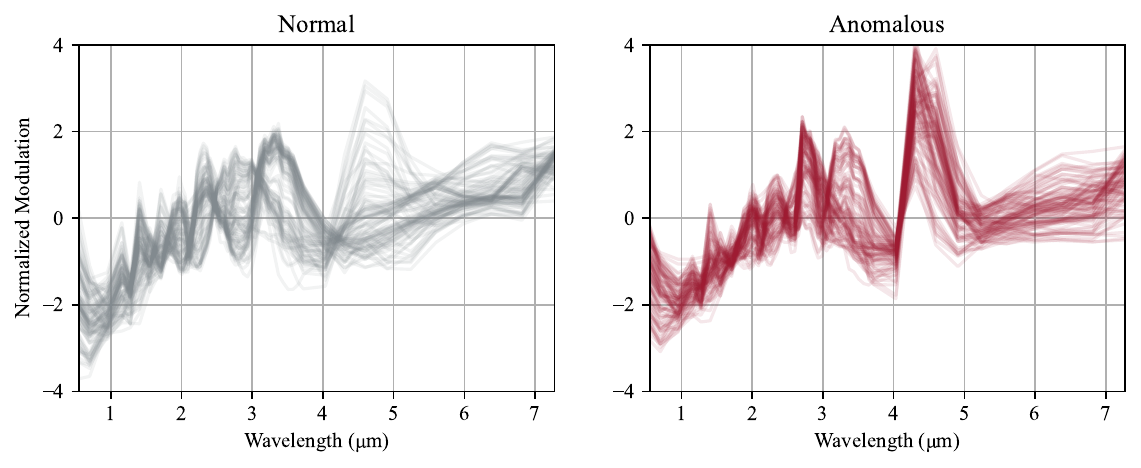}
    \caption{Comparison of normal spectra (left panel) and anomalous spectra (right panel). For better readability, here we show only 100 normalized spectra of each population. }
    \label{fig:spectra-comparison}
\end{figure}

To perform anomaly detection, we begin by constructing two datasets, one of ``normal'' and one of ``anomalous'' spectra. 
To simulate a novelty detection scenario, we choose to label carbon dioxide (CO$_2$) rich atmospheres as anomalous. Specifically, we identify a spectrum as anomalous if the relative mixing ratio of CO$_2$ exceeds 5\%. The {\em relative} mixing ratio $r_i$ is defined in terms of the individual concentrations $X_i$ as follows \citep{2022ApJ...939...95M} 
\begin{equation}
r_i \equiv \frac{X_i}{\sum_j X_j},
\label{eq:rdef}
\end{equation}
where the sum in the denominator runs over all absorbers listed in Table~\ref{tab:abundances}.
Conversely, a spectrum is considered to be normal if the relative mixing ratio of CO$_2$ is below $10^{-4}$. This choice enables us to create a clearly defined chemical anomaly while maintaining physical plausibility.  Although carbon dioxide is technically still present in the normal class, it is effectively undetectable in transmission, overwhelmed by the stronger features of other absorbers. Note that the results from \cite{Forestano_2023} indicated that out of the four absorbers, H$_2$O, CH$_4$, CO$_2$, and NH$_3$, the case of CO$_2$ is the most challenging in terms of anomaly detection and this is why in this paper we only focus on this most difficult case. 

The two populations are illustrated in Figure \ref{fig:normal_vs_anomalous}. The left panel shows a scatter plot of the log concentration of CO$_2$ versus the temperature.  The gray points represent the normal population (CO$_2$-poor) and the red points represent the anomalous population (CO$_2$-rich). In the right panel of Figure \ref{fig:normal_vs_anomalous} we also show histogrammed distributions of the two populations versus the log concentration of CO$_2$. We see that the two populations are well separated which indicates that the two classes, normal and anomalous, are unambiguously defined.

We simulate realistic observational conditions by adding uncorrelated Gaussian noise at various levels (0, 10, 20, 30, and 50~ppm). We then normalize each spectrum individually (sample-wise) to have zero mean and unit variance. 
This normalization step removes absolute scale information and is justified by the fact that the values for the mean and standard deviation of a given spectrum are largely independent of the planet's atmospheric composition \citep{2022ApJ...939...95M, Forestano_2023}. We show some typical examples of normal and anomalous spectra (after their normalization) in Figure \ref{fig:spectra-comparison}. After the normalization step, we can clearly see well-defined CO$_2$ features in the right panel of Figure \ref{fig:spectra-comparison} --- for example, between 4 and 5 microns.

\section{Methods} \label{sec:methods}

\subsection{Dimensionality reduction methods}

\begin{figure}[t]
    \centering
    \includegraphics[width=\textwidth]{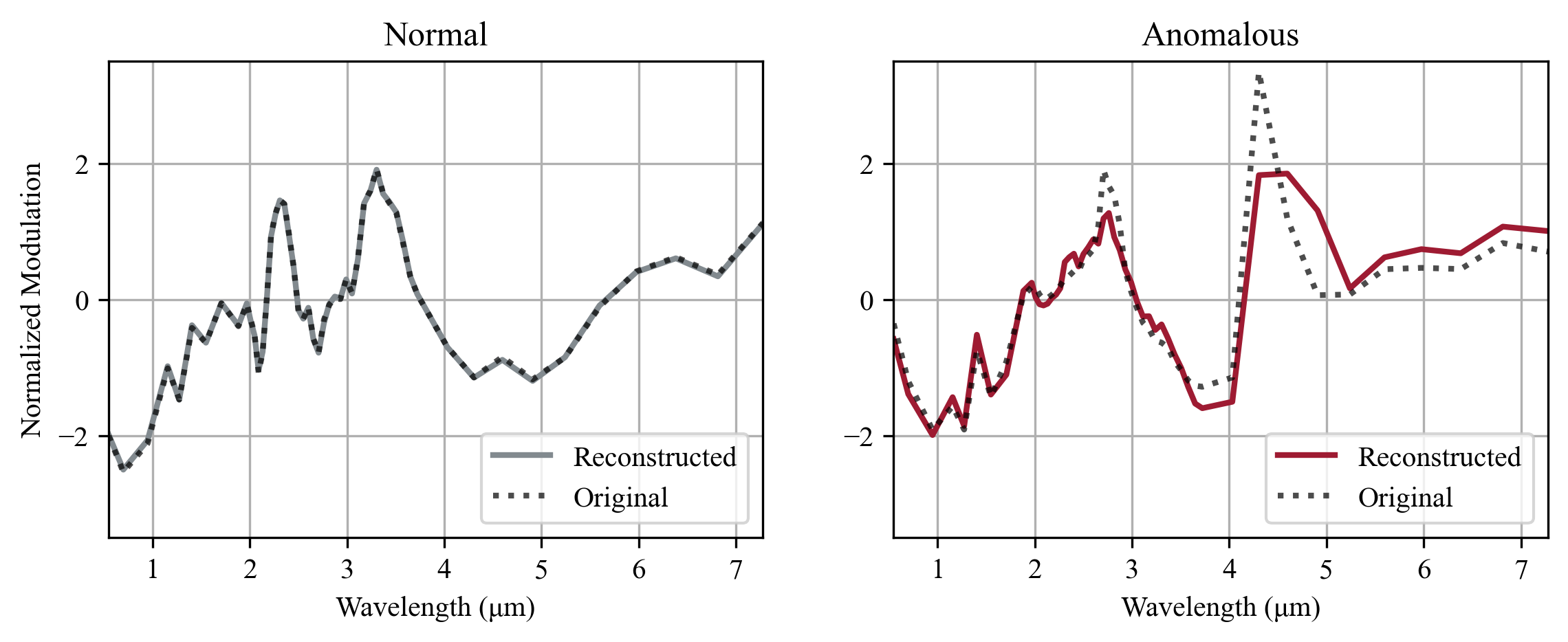}
    \caption{Reconstruction comparison for normal (left) and anomalous (right) spectra (after normalization). For normal data, the autoencoder closely matches the original spectrum. For anomalous data, the reconstruction fails to capture key features, showing larger differences. The spectra pictured here are `ideal' and have not had any noise added.}
    \label{fig:reconstruction_example}
\end{figure}

There are a number of dimensionality reduction techniques that are used for compressing the data into a lower-dimensional feature space. In our study, we use an autoencoder architecture, which has a dual advantage --- it not only provides a near lossless compression, but can also be used as an anomaly detection tool (see Section~\ref{sec:ADmethods}). Autoencoders are a class of neural networks in which data is encoded into a lower-dimensional representation and then decoded back into the original data space. In the simplest version, the network is trained to make the output match the input as closely as possible by penalizing deviations. Specifically, the loss function is the \emph{reconstruction loss}:
\begin{equation}
\mathscr{L}(\vec{x}) = \mathbb{E}_{\vec{x} \in V} \|\vec{x} - D(E(\vec{x}))\|^2,
\label{eq:recoloss}
\end{equation}
which quantifies the average mismatch between the input $\vec{x}$ and the reconstructed output $D(E(\vec{x}))$. Here $E(\vec{x})$ is the encoded representation in the latent space, $D$ is the decoder, $\vec{x}$ is a transit spectrum drawn from the space $V$ of allowed physically meaningful spectra. Note that in applications to novelty detection, the expectation value in (\ref{eq:recoloss}) is computed over training data taken from the normal population in $V$. In other words, the autoencoder never sees any anomalous examples during training which is why the quality of the reconstruction of an anomalous spectrum is significantly worse. Figure \ref{fig:reconstruction_example} illustrates this point by showing a reconstruction for a representative normal spectrum (left panel) and a typical anomalous spectrum (right panel). In each case, the dotted line shows the original spectrum, while the solid line depicts the reconstructed spectrum. As expected, the reconstruction for the normal spectrum in the left panel is nearly perfect, while the two spectra shown in the right panel are noticeably different. This suggests that the reconstruction error can be used as a proxy for novelty: spectra that the model fails to reconstruct accurately would be flagged as anomalous (see Section~\ref{sec:ADmethods} below).

\begin{figure}[t]
    \centering
    \includegraphics[width=0.7\linewidth]{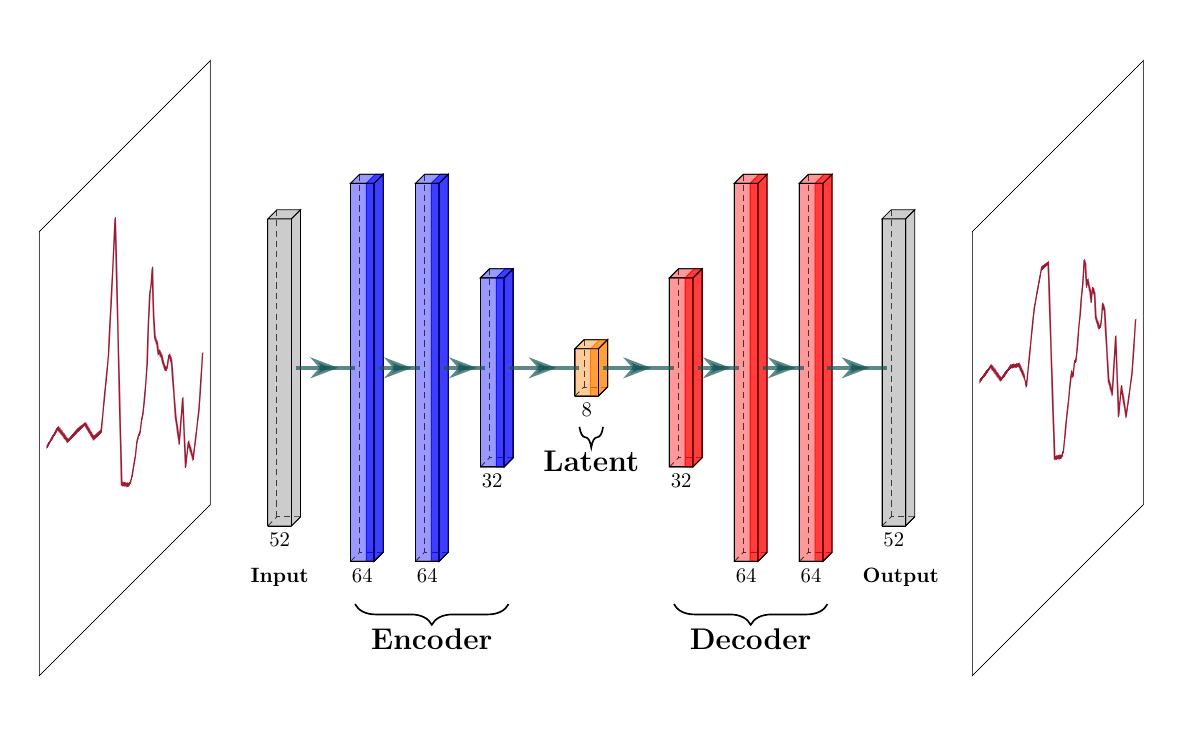}
    \caption{A visualization of the architecture of the autoencoder network. Each colored block represents a layer of the network and the shaded band indicates a ReLU activation function, present on all layers except the output.}
    \label{fig:network-architecture}
\end{figure}

We implement a fully connected autoencoder in PyTorch. The encoder compresses the 52-dimensional input spectra through hidden layers of sizes 64, 64, and 32 into an $8$-dimensional latent space, using LeakyReLU activations (see Figure~\ref{fig:network-architecture}).
The decoder reverses this architecture to reconstruct the input spectrum from its latent representation. The model is trained using Adam optimizer and mean squared error loss on the `normal' (low-CO$_2$) spectra, with Gaussian noise added to simulate observational uncertainty. We also save the latent representation of the data for later use with the other anomaly detection methods discussed in the next subsection. Further details of our implementation can be found in our public code available on Github.

\subsection{Anomaly detection methods}
\label{sec:ADmethods}

In the analysis presented in the next section, we consider several classical anomaly detection methods:

\begin{itemize}
    \item \textbf{Reconstruction Error:} 
    The learned lower-dimensional latent representation captures the most salient features of the data. However, the learned mapping tends to be faithful only for data drawn from the training distribution. In other words, the autoencoder becomes specialized in reconstructing the training data and may generalize poorly to inputs outside the distribution from which the training data was drawn. Figure \ref{fig:reconstruction_example} is a typical example of this: the reconstructed spectrum is nearly indistinguishable in the case of normal data (left panel) but differs noticeably in the case of anomalous data (right panel). This makes the reconstruction loss (\ref{eq:recoloss}) a natural test statistic for novelty detection: a high reconstruction error can serve as a signal of unfamiliarity. 
    
    

    \item \textbf{One-Class Support Vector Machine (1-Class SVM):} One-Class SVM is a boundary-based anomaly detection algorithm that learns a decision surface separating the bulk of the data from the origin in feature space \citep{vapnik95}. Unlike standard SVMs, which require labeled data from multiple classes, the 1-Class SVM is trained only on normal data and attempts to encapsulate the region where most of the training data lies. Any points falling outside this learned boundary at test time are flagged as anomalies. Conceptually, the method tries to find a hyperplane that maximizes the margin between the origin and the training data while allowing a small fraction of the data to fall outside the boundary as potential outliers.

    We employ the radial basis function (RBF) kernel to allow for nonlinear boundaries, which is particularly important given the complexity of both the spectral and latent data manifolds. The hyperparameter \texttt{gamma} ($\gamma$) controls the width of the RBF kernel: smaller values lead to smoother, more global boundaries, while larger values allow for tighter fitting to local variations. We set $\gamma = 0.2$ to balance flexibility with generalization. The parameter \texttt{nu} ($\nu$) controls the trade-off between the fraction of allowed training errors (i.e., margin violations) and the complexity of the decision boundary; effectively, it can be interpreted as an upper bound on the expected proportion of anomalies in the training data. We use $\nu = 0.01$, which corresponds to assuming that approximately 1\% of the training data may be atypical or noisy. These hyperparameters were chosen to reflect our prior belief that the training data consists almost entirely of normal samples, while still allowing some tolerance for minor deviations.
    
    \item \textbf{K-means Clustering:} K-means clustering for anomaly detection is an efficient implementation of a simpler idea: using the minimum distance to any point in the normal dataset as a test statistic. Having a small distance from a point which is known to be normal should indicate normality, and, assuming the dataset is representative of the distribution of normal points, any point having a large minimum distance should be considered anomalous. The drawback to this approach is the complexity of computing the minimum distance: the naïve algorithm has complexity $O(N_\text{dataset})$ at inference time, and improved algorithms such as K-D Trees still require $O(\log(N_\text{dataset}))$ at inference, this would be expensive considering that all other methods we consider run in constant time. One solution is to avoid computing the distances to every point in the dataset by finding a smaller set of representative points which adequately capture the shape of the distribution. K-means clustering is employed to do just that, it produces a set of cluster centers or `prototype' points, and then for any new query points we need only compute the distance with these $K$ points, the minimum of which is used as the anomaly score. In this work we used the default value of the hyperparameter $K=10$ in {\tt scikit-learn} \citep{scikit-learn}.
    
    \item \textbf{Local Outlier Factor (LOF):} Local Outlier Factor is a density-based anomaly detection algorithm which, unlike distance-based or global methods, is designed to detect anomalies by comparing the local density of a given point to the densities of its neighbors \citep{Breunig2000}. The core intuition is that normal data points tend to reside in regions of approximately uniform density, while anomalies are located in regions of significantly lower density relative to their neighbors. Specifically, LOF computes the local reachability density (LRD) of each data point, which is a measure of how far away its neighbors are, adjusted for the density of those neighbors themselves. The anomaly score is then defined as the ratio of the average LRD of the point’s neighbors to the LRD of the point itself; points with much smaller density than their neighbors will thus receive high LOF scores and be flagged as outliers.

    Because LOF is sensitive to local structure, it can be especially effective in cases where the data manifold is highly non-uniform or contains clusters of varying density, which may arise naturally in both the original spectral data and in the latent spaces learned by the autoencoder. In this work, we apply LOF to both spaces, recognizing that the learned latent space may concentrate the essential features of the data into lower-dimensional manifolds where density variations become even more informative for anomaly detection. The number of neighbors used to compute local densities is a key hyperparameter that controls the scale of locality; here we use $n_\text{neighbors} = 20$.
    
\end{itemize}

The last three methods will be applied in both the original spectral space and in the learned latent space, allowing for a comparison of the two approaches. We compare these methods using Receiver Operating Characteristic (ROC) curves and Area Under the Curve (AUC) metrics across different noise levels. In all cases, the models are trained exclusively on ``normal'' spectra and evaluated on a held-out mixture of normal and anomalous samples.

\begin{table}[t]
\centering
\caption{AUC scores for each anomaly detection method, for each noise level and feature space. For each noise level, the overall highest score among the seven tested methods is denoted in bold.}
\begin{tabular}{clcccc}
\toprule
\textbf{Noise (ppm)} & \textbf{Feature Space} & \textbf{Reconstruction Loss} & \textbf{1-Class SVM} & \textbf{K-means} & \textbf{LOF} \\
\midrule
\multirow{2}{*}{10}  & Spectral       & 0.9978 & 0.9731 & 0.9808 & 0.9960 \\
\cmidrule(lr){2-6}
                     & Latent         & -  & 0.9994 & 0.9956 & \textbf{0.9995} \\
\midrule
\multirow{2}{*}{20}  & Spectral       & 0.9565 &0.9501 & 0.9724 & 0.9411 \\
\cmidrule(lr){2-6}
                     & Latent         & - & 0.9894 & 0.9921 & \textbf{0.9940} \\
\midrule
\multirow{2}{*}{30}  & Spectral       & 0.8925   &  0.9132 &  0.9482  & 0.8418 \\
\cmidrule(lr){2-6}
                     & Latent         & - & 0.9641 & \textbf{0.9824} & 0.9785 \\
\midrule
\multirow{2}{*}{50}  & Spectral       & 0.7910  & 0.8241 & 0.8835 & 0.6598 \\
\cmidrule(lr){2-6}
                     & Latent         & - &  0.8781 & \textbf{0.9536} & 0.9155 \\
\bottomrule
\end{tabular}
\label{tab:auc_scores}
\end{table}

\section{Results} \label{sec:results}

\begin{figure*}[t]
    \centering
    \includegraphics[width=0.8\textwidth]{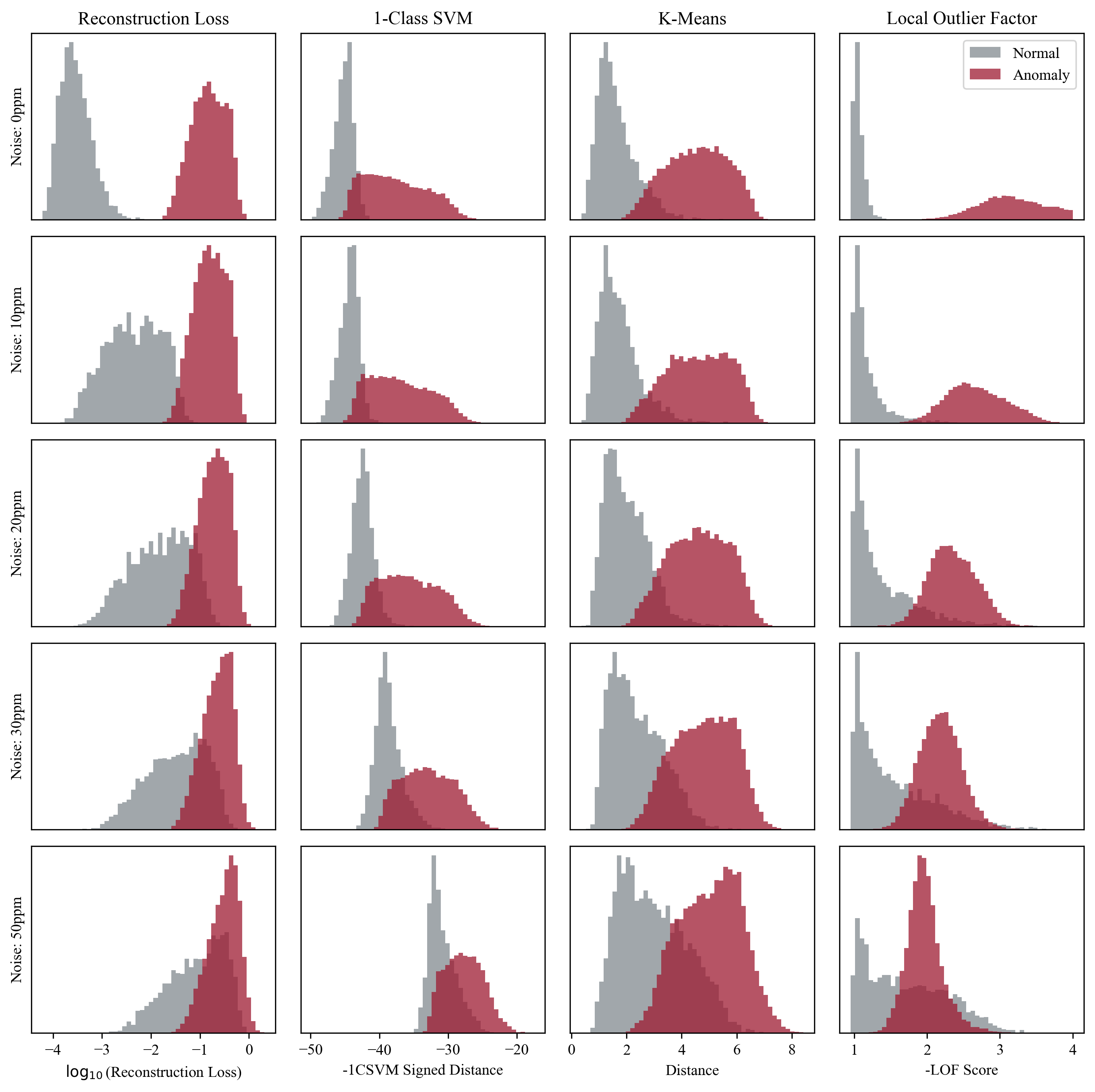}
    \caption{Histograms of the corresponding anomaly scores for each method in spectral space. Normal spectra are represented in gray and anomalous spectra are represented in red. Both populations show some degree of overlap, especially with a high level of noise. The first column shows histograms of the reconstruction loss as defined in (\ref{eq:recoloss}).
    }
    \label{hists_spec}
\end{figure*}

\begin{figure*}[phtb]
    \centering
    \includegraphics[width=0.8\textwidth]{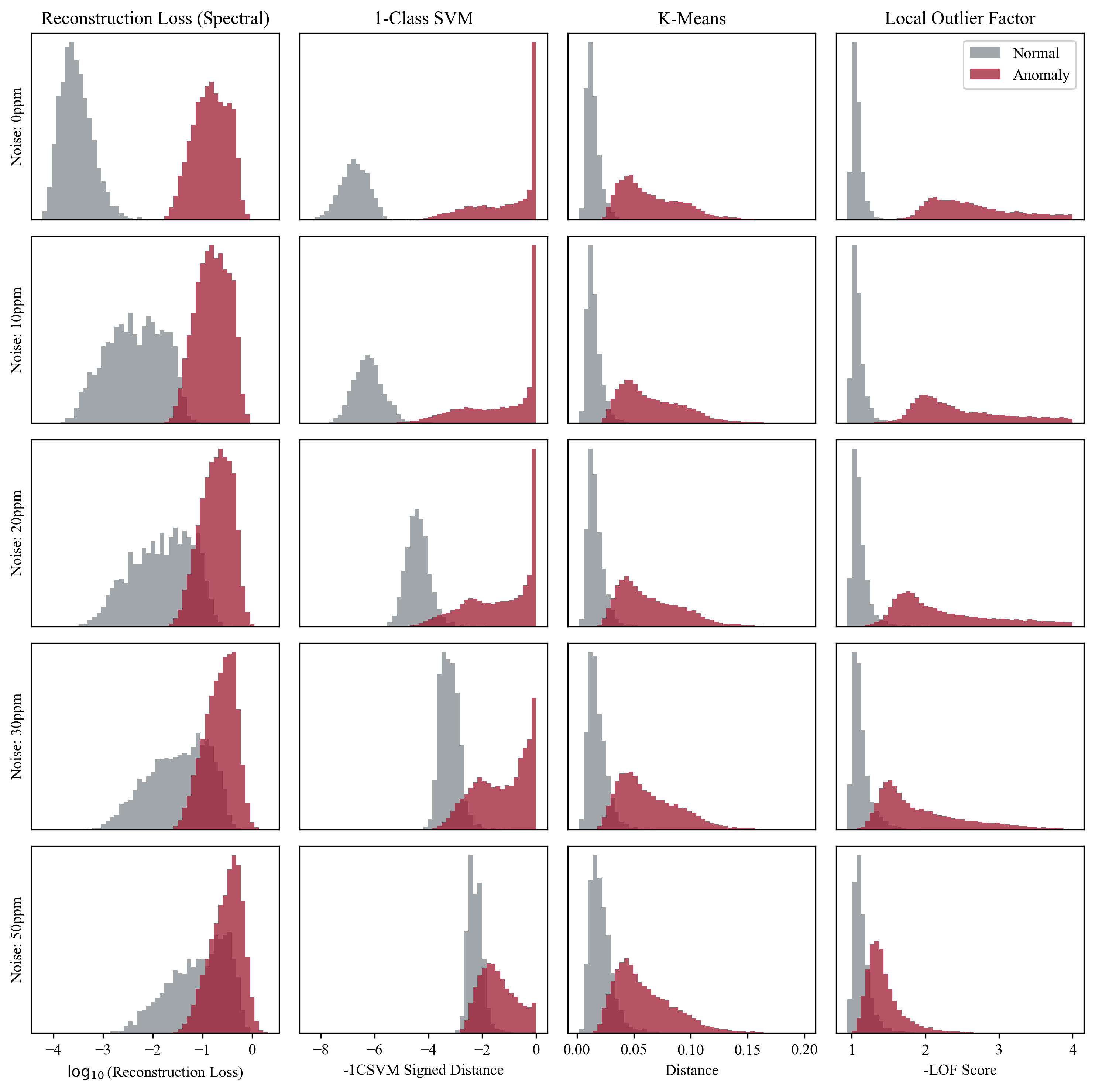}
    \caption{The same as Figure~\ref{hists_spec}, but when each method is applied in the latent space.
    Note that the first column is the same as the first column in Figure \ref{hists_spec}.}
    \label{hists_lat}
\end{figure*}

\begin{figure}[htbp]
    \centering

    \begin{minipage}[t]{0.48\textwidth}
        \centering
        \includegraphics[width=\linewidth]{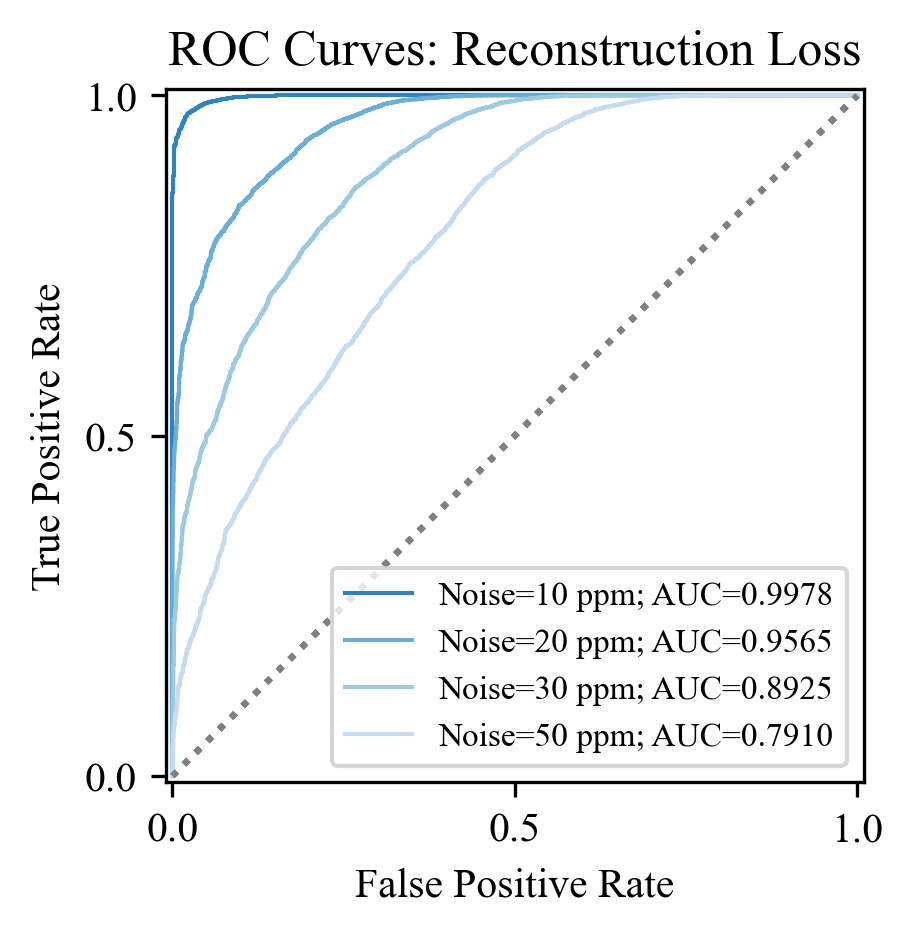}
        \caption{ROC curves for the reconstruction loss method with noise levels applied at 10, 20, 30, and 50 ppm. 
        The resulting AUC is quoted in the legend.
        A higher AUC value indicates better overall performance.}
        \label{fig:roc_reconstruction}
    \end{minipage}%
    \hfill
    \begin{minipage}[t]{0.48\textwidth}
        \centering
        \includegraphics[width=\linewidth]{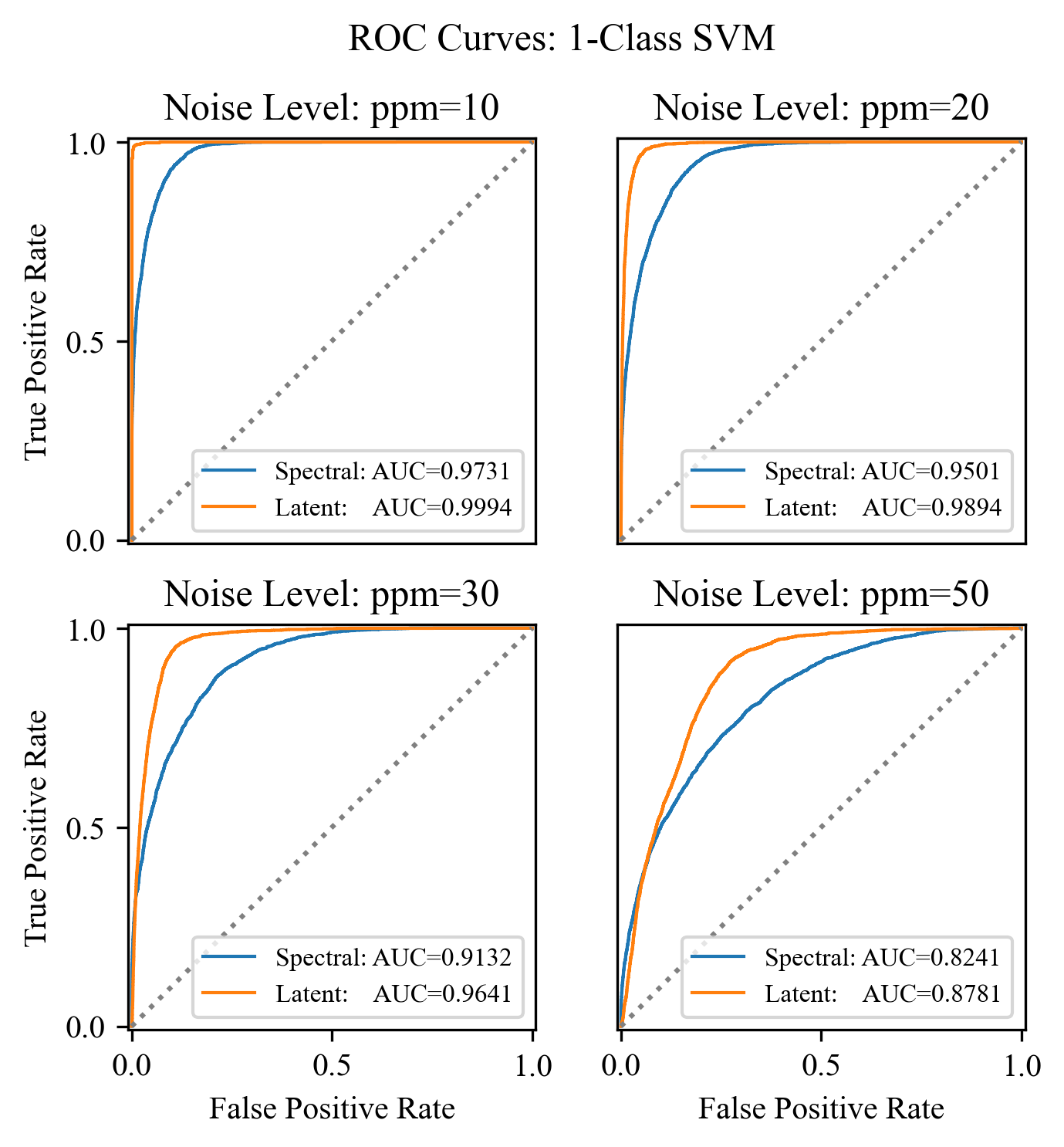}
        \caption{ROC curves for the 1-class SVM method with noise levels applied at 10, 20, 30, and 50 ppm. Orange (blue) curves correspond to the case when the method is applied in the latent (spectral) feature space.
        }
        \label{fig:roc_svm}
    \end{minipage}

    \vspace{0.5cm} 

    \begin{minipage}[t]{0.48\textwidth}
        \centering
        \includegraphics[width=\linewidth]{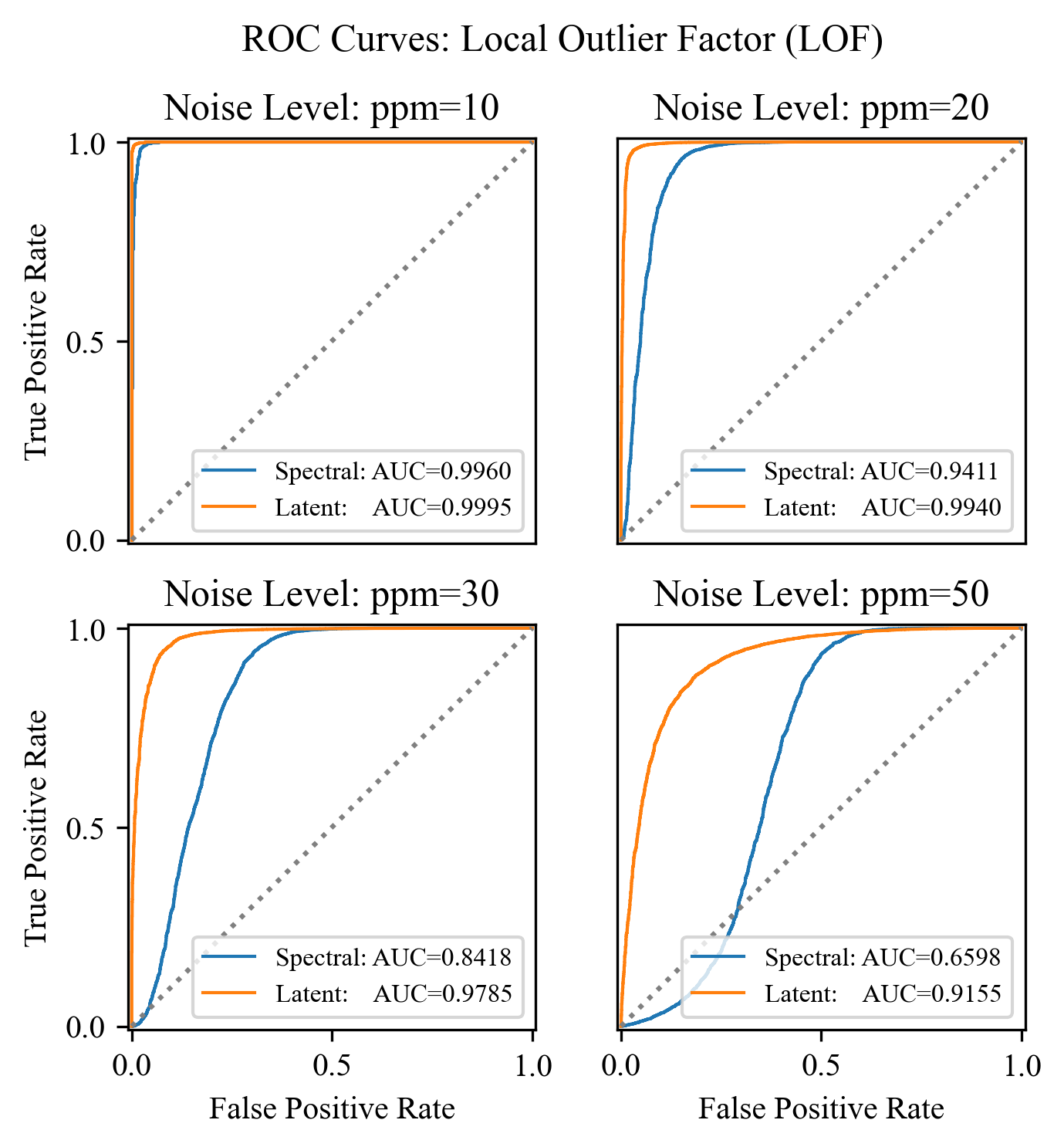}
        \caption{The same as Figure~\ref{fig:roc_svm}, but for the LOF method.
        }
        \label{fig:roc_lof}
    \end{minipage}%
    \hfill
    \begin{minipage}[t]{0.48\textwidth}
        \centering
        \includegraphics[width=\linewidth]{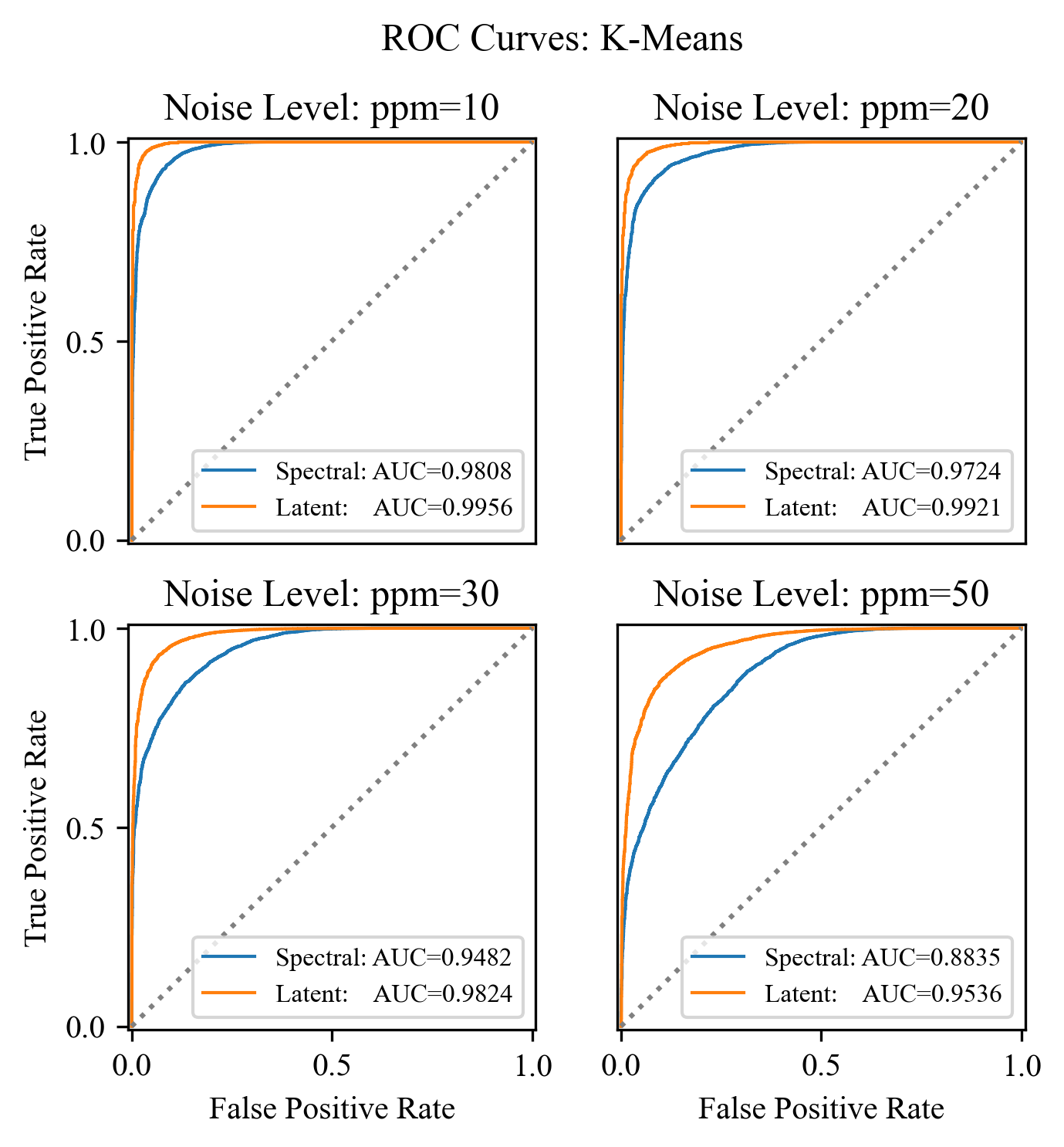}
        \caption{The same as Figure~\ref{fig:roc_svm}, but for the K-means method.
        }
        \label{fig:roc_kmeans}
    \end{minipage}

\end{figure}

\begin{figure*}[ht]
    \centering
    \includegraphics[width=0.8\textwidth]{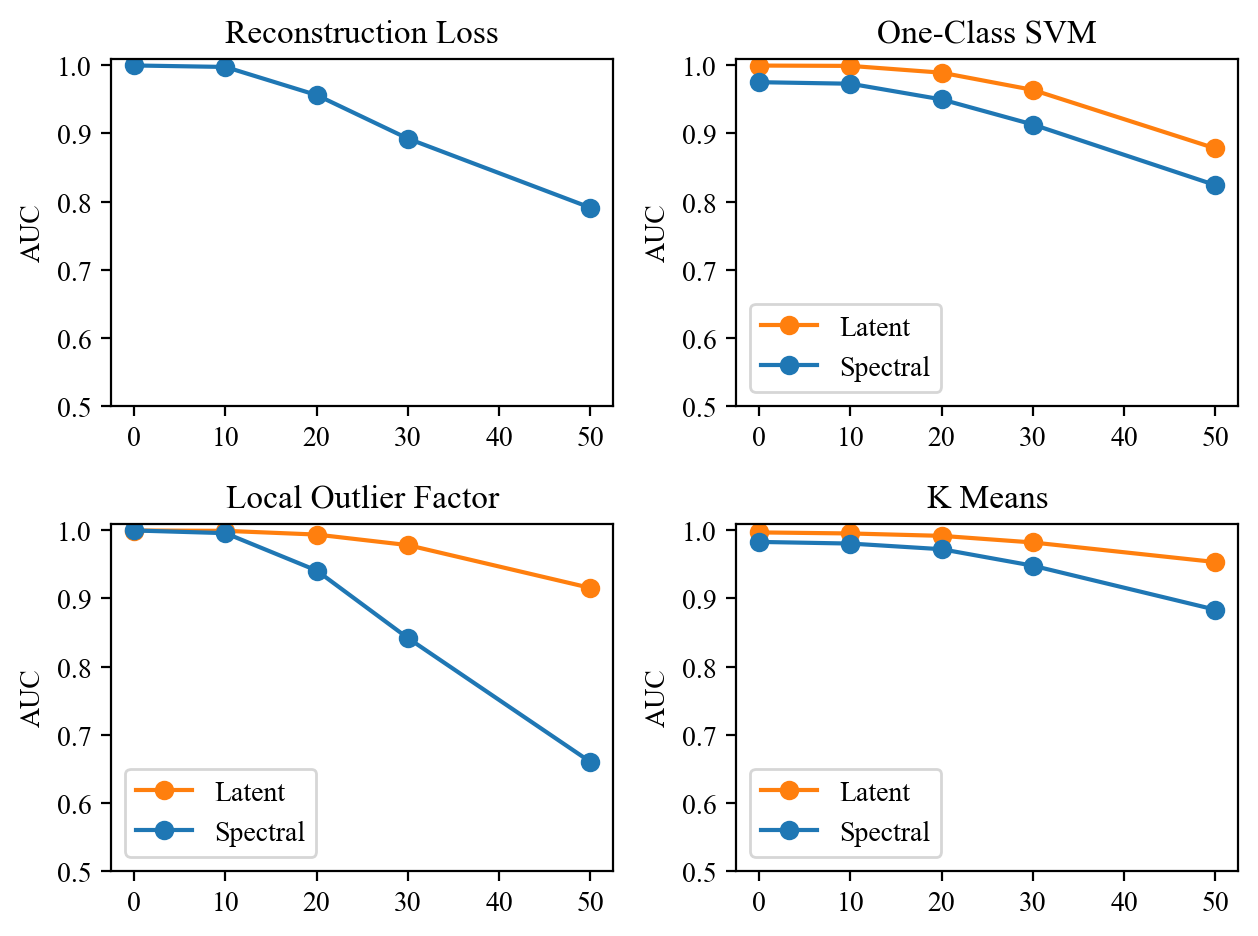}
    \caption{Comparison of AUC (Area Under the Curve) values across different methods, spaces, and noise levels, as indicated in the legends and titles. 
    While performance generally decreases with increasing noise, the methods applied in the latent space consistently outperform the methods applied in the original spectral space.}
    \label{money_plot}
\end{figure*}

The results from our numerical experiments are collected in Table~\ref{tab:auc_scores} and Figures~\ref{hists_spec}-\ref{money_plot}. We explore different aspects of the anomaly detection procedure. First, we investigate the effect of using a latent representation of the spectral data --- each of the three anomaly detection methods was applied in both the original spectral space (Figure~\ref{hists_spec}) and in the low-dimensional latent space (Figure~\ref{hists_lat}). We also monitor the reconstruction loss (first columns in Figure~\ref{hists_spec} and Figure~\ref{hists_lat}, as well as Figure~\ref{fig:roc_reconstruction}) as a useful benchmark. Second, we compare the performance of the different methods relative to each other, for a given noise level (the rows in Table~\ref{tab:auc_scores} and in Figures~\ref{hists_spec} and \ref{hists_lat}). Third, we track the performance of the methods for increasing levels of noise (Figures~\ref{fig:roc_reconstruction}-\ref{fig:roc_kmeans}). Figure~\ref{money_plot} provides a concise summary of the results across different feature spaces, anomaly detection methods, and noise levels.

\subsection{Performance in different spaces}

The anomaly detection performance was studied in two different spaces: the spectral space and the latent space. We found that anomaly detection works better in the latent space. All models presented in this study show higher AUC values in latent space compared to spectral  space. For example, the top performing model at each noise level in Table~\ref{tab:auc_scores} was applied in latent space. 
The advantage of the use of latent space is further demonstrated in Figures~\ref{fig:roc_svm}, \ref{fig:roc_lof}, and \ref{fig:roc_kmeans}, which show ROC curve\footnote{A receiver operating characteristic (ROC) curve is a plot that illustrates the performance of a binary classifier. The true (false) positive rate is plotted on the $y$ ($x$) axis and the curve is built by continuously varying the discrimination threshold of the classifier. An ideal ROC curve would pass through the point $(0,1)$ in the upper left corner, which corresponds to a perfect classifier with no errors.} comparisons between spectral and latent spaces for the 1-class SVM method, the Local Outlier Factor method and the K-means method, respectively. In each panel, the orange (blue) curve corresponds to using the latent (spectral) feature space. In every case, the orange curve is above the blue curve, indicating a better discriminating power. The AUC values from the numerical experiments shown in Figures~\ref{fig:roc_svm}, \ref{fig:roc_lof}, and \ref{fig:roc_kmeans}, are collected in Table \ref{tab:auc_scores} and plotted in Figure \ref{money_plot} as a function of the noise level.

\subsection{Effect of noise}

We added different levels of gaussian noise to the spectra in our database: 10, 20, 30, and 50 ppm. This lets us study how noise affects the performance of our anomaly detection methods and simulates the effect of real observational uncertainty. 
Results show that the performance is stable up to 30 ppm, with only small decreases in the area under the ROC curve (AUC), staying around 0.9 for almost every considered case. This can be seen in the ROC curves presented in Figures \ref{fig:roc_reconstruction}, \ref{fig:roc_svm}, \ref{fig:roc_lof}, and \ref{fig:roc_kmeans}. At 50 ppm, the separation between normal and anomalous spectra becomes less clear and the models had a harder time separating normal and anomalous spectra. This trend is visually illustrated in Figures~\ref{hists_spec} and \ref{hists_lat}, where the increased level of noise leads to a larger overlap between the normal and anomalous populations.
The overlap was greater in spectral space, where the noise directly affects the input features. Figure \ref{money_plot} and Table \ref{tab:auc_scores} summarize this behavior.  
Our results show that the use of latent space is more robust to noise than the spectral space. This makes latent space a good choice for representing real observational data (which always includes noise) for subsequent analysis. It also highlights the necessity of minimizing the observational error to begin with.

\subsection{Method comparison}

In this study, we tested four main anomaly detection methods: reconstruction loss, 1-class SVM, K-means clustering, and Local Outlier Factor. Each method was tested in both spectral space and latent space. 
No single method was the best under all circumstances, which is consistent with the general intuition developed in the machine learning community. Overall, the K-means method in latent space showed consistently good performance, as well as robustness with respect to noise.
The corresponding ROC curves are showed in Figure \ref{fig:roc_kmeans}. 
The 1-class SVM also performed well, but with a little bit more sensitivity to the noise level (Figure~\ref{fig:roc_lof}). 
The reconstruction loss method, which is very intuitive and simple to use and implement, was very effective at low noise, but quickly deteriorated as the noise increased. 
The LOF method had mixed results. It worked exceptionally well at low noise levels (10 and 20 ppm) in latent space, but its performance deteriorated quickly at higher noise levels in spectral space, where at 50 ppm, it gave the worst result with AUC below 0.7. At the same, LOF peformed well in latent space at all noise levels.


\section{Conclusions and discussion} \label{sec:conclusions}


In this paper, we studied anomaly detection in exoplanet transmission spectra using an autoencoder framework for dimensionality reduction to a latent space. We focused on the detection of CO$_2$ as an example of an anomalous chemical signature. Our findings demonstrate that anomaly detection is most effective when performed within the latent space, a result that holds consistent across various detection methods and noise levels. Among the evaluated anomaly detection techniques, K-means clustering in the latent space yielded superior performance for distinguishing between normal and anomalous spectra. Furthermore, we demonstrate that this approach is robust to noise levels up to 30 ppm (consistent with realistic space-based observations) and remains viable even at 50 ppm when leveraging latent space representations.



This study highlights the usefulness of latent space as a robust diagnostic tool. By compressing spectra into a lower-dimensional manifold, the autoencoder preserves essential physical features. Consequently, normal spectra converge into tight clusters within the latent space, whereas anomalous signatures emerge as distinct outliers, making them easier to detect with distance based methods like K-means.


This methodology can be used to automatically detect unusual chemical patterns in exoplanet atmospheres, which would be useful for large planetary surveys like Ariel. The Ariel mission's goal is to observe one thousand exoplanet atmospheres through transmission spectroscopy. Manual inspection or full atmospheric retrievals for all targets will be computationally expensive. The approach presented here can be used as a fast and automated pre-screening tool. It can help identify unusual spectra due to, e.g., unexpected molecules, that deserve further detailed analysis or follow-up observations. Since the anomaly detection methods are trained only on normal data, they do not require labeled anomalies and can generalize well to potential surprises.


This study also has some limitations. We focused on a single type of anomaly, namely CO$_2$-rich atmospheres, and the ABC database that we used was generated with a few simplifying atmospheric assumptions. Real observations may include clouds, hazes, stellar contamination, and instrumental systematics. Applying this method to real data will require careful preprocessing and training on more diverse simulations. It may also require extending the training dataset to include more molecules or equilibrium chemistry models.


Overall, this work shows that combining dimensionality reduction with anomaly detection is a powerful approach. It is well suited for the analysis of large  datasets and represents a promising tool for future space missions.

\begin{acknowledgments}
The work of AR, EP and KTM is supported in part by the Shelby Endowment for Distinguished Faculty at the University of Alabama. 
\end{acknowledgments}

\bibliography{references}{}
\bibliographystyle{aasjournalv7}


\end{document}